\documentstyle[12pt]{article}

\author{Ludger Hannibal\\
Fachbereich Physik, Carl v. Ossietzky Universit\"at Oldenburg\\
D-26111 Oldenburg, Germany\\
e-mail: hannibal@caesar.physik.uni-oldenburg.de}
\title{Neutral Kaon decay without symmetry violation: A
nonstandard  theoretical
speculation
}
\date{November 28, 1995}

\begin{document}

\maketitle
\begin{abstract}
It is shown that if antiparticles are realized in quantum field theory
 by negative frequency states, which  nevertheless have
 positive energy density, the resulting
theory   provides a qualitative explanation for the experiments on the
neutral K mesons, without assumming any symmetry violation.
\end{abstract}

\section{Introduction}

The existence of the positron was predicted on the basis of the negative
energy solutions of the Dirac equation, but then a physical interpretation of
negative
energy states was not found in the framework of single-particle theory.
So, in the usual construction of quantum field theory antiparticle states
have positive frequencies, and we have an axiomatic positivity condition on
the free antiparticle spectrum \cite{Bog,Streater,ReedSimon}.

But there is also an alternative.
In 1941 Stueckelberg \cite{Stueckelberg} gave a
classical picture of pair creation and annihilation where the
electron-positron pair is described by a single world line which is
reflected in time by electromagnetic interaction. The electron
part of the world line has $dt/d\tau >0$, where $\tau $ is the proper time,
the positron part has $dt/d\tau <0$. For particles the classical energy
variable $p_0$ is positive, for antiparticles negative. Stueckelberg also
devised a quantum theory which is in correspondence to this classical
picture by introducing the proper time as an evolution parameter into
quantum theory, with the mass as the conjugate invariant quantity, masses
 become the
eigenvalues of the operator $i\partial_{\tau} $.
 Particles are decribed by positive frequency solutions,
antiparticles have negative frequencies. It is well known \cite{Fanchi} that
this manifest covariant quantum theory solves the problems of localization
and zitterbewegung which arise when we consider only the time-development
in a $3\times 1$ decomposition.

But Stueckelberg's classical and quantum picture, further extended by
 Feynman \cite{Feynman} to spin 1/2, could not solve
the negative energy problem. But there exists a solution
 to this problem, in the way that,
 negative frequency states  are
interpreted in  a consistent way with positive energy density,
with corresponding pictures in classical theory, quantum mechanics and
quantum field theory. This theory is used to analyse neutral K-meson decay
with the result that no symmetry violation needs be assumed in order to expain
the experimental observations.

\section{Positivity of the energy density}

We start with classical mechanics \cite{Stueckelberg}. The constraint
\begin{equation}
\label{1}\dot x^\mu (\tau )g_{\mu \nu }(x(\tau ))\dot x^\nu (\tau )=c^2
\end{equation}
on the trajectories of classical point particles allows for solutions with
negative $\dot x^0=cdt/d\tau $, i.e. $t=-\tau $ for a particle at rest.
These solutions were usually discarded, postulating $t=\tau $ for a particle
at rest, until Stueckelberg showed that these solutions may be interpreted
as antiparticles. With respect to the physical interpretation it is
important to note that the solutions of the canonical equations with $%
dt/d\tau <0$, which also have negative canonical energy $p_0<0$, possess a
positive energy density \cite{Hannibal1}
\begin{equation}
\label{2}T_0^0=\int d\tau p_0\dot x^0\delta ^4(x-x(\tau )),
\end{equation}
which is positive due to the two negative signs of $p_0$ and $\dot x^0$. (We
use the sign convention $+---$ for the metric.)

With respect to quantum theory,
Klein-Gordon theory has an energy density which is positive
irrespective of the sign of the eigenvalues of the  operator
$\hat p_0=\frac{i\hbar }c\frac \partial {\partial t}$. A solution to the
positivity problem for spin 1/2 was first indicated by Arshansky and Horwitz
\cite{Arshansky}. The argument is from representation theory.
Since the energy density operators \cite{Hannibal2}
\begin{equation}
\label{3}\hat T_0^0=\hat v^0\hat p_0,\quad \hat p_0=\displaystyle
\frac{i\hbar }c\displaystyle \frac \partial {\partial t},\quad \hat
v^0=\left\{
\begin{array}{cc}
\frac{i\hbar }{mc}\frac \partial {\partial t} & {\rm for\; spin 0} \\ \gamma
^0 & {\rm for\; spin 1/2}
\end{array}
\right.
\end{equation}
always have a positive spectrum
 the question of a positive energy density is a matter of
choosing a positive definite scalar product which preserves the sign of the
eigenvalues when expectation values are taken. The idea by Arshansky and
Horwitz \cite{Arshansky} was to start with two-component spinors which
transform under Wigner's unitary induced representation or its conjugate
representation, for which we have a manifest positive norm \cite{Wigner}.
Then the embedding of the two-component spinors into a space of
four-component spinors is done in a norm-preserving way, leading to the
positive norm
\begin{equation}
\label{8}N=\int \frac{d^3p}{p_0}\psi ^{*}\gamma ^0\psi
\end{equation}
on the embedded subspace \cite{Hannibal5}.
 We note that the functional form
(\ref{8}) of the norm, with a sign included for antiparticles, is
the same as in standard quantum field theory, where
antiparticle states are realized with positive frequencies \cite{Bog}, p.
188.

\section{Second Quantization, Discrete Symmetries}

The second quantization of negative frequency states can be carried out just
as that for positive frequency states, with the antiparticle sector related
to the particle sector by an antilinear, antiunitary charge conjugation
transformation, which is precisely the one from usual Dirac and Klein-Gordon
theory extended to Fock space \cite{Hannibal3,Hannibal4}. The decisive
point is the construction of the energy density operator, which is done in
the following different way compared to standard theory. In standard theory
the free Dirac quantum field $\hat \psi (x^i)$ is constructed as a time-zero
field with Hermitian conjugate $\hat \psi ^{\dagger }(x^i)$, so that $\hat
\psi ^{\dagger }\hat \psi $ is a positive operator. The time dependence is
generated in a Heisenberg picture by $\hat \psi (x^i,t)=e^{iHt}\hat \psi
(x^i)e^{-iHt}$. The energy density operator $:\hat \psi ^{\dagger }i\partial
_t\hat \psi +h.c.:$ then is positive definite if and only if the Hamiltonian
is positive. In the $4\times 1$ decomposition the adjoint $\overline{\hat
\Psi }(x^\mu )$ of the field $\hat \Psi (x^\mu )$ is constructed with
respect to the four-dimensional, Lorentz invariant scalar product on Fock
space, such that $\overline{\hat \Psi }\hat \Psi $ is positive, and the
dependence on proper time is generated by a positive definite mass operator $%
M$, $\hat \Psi (x^\mu ,\tau )=e^{iM\tau }\hat \Psi (x^\mu )e^{-iM\tau }$.
The energy density operator is given by $\overline{\hat \Psi }\gamma
^0i\partial _t\hat \Psi $ and now the negative entries of the matrix $\gamma
^0$ allow for positive definiteness only if the operator $\hat p_o=H$,
generator of inifinitesimal translations in time is negative on antiparticle
states. This construction was carried out rigorously \cite{Hannibal4}. The
use of negative frequencies is not in contradiction to the axiomatic
approach, since it is well known that positivity is independent of all other
axioms \cite{Streater,ReedSimon}.

In  original Dirac and Klein-Gordon theories the charge conjugation
 is an antilinear transformation.
 In the parametrized theory this charge conjugation
transformation is extended to parameter-dependent wave-equations, whereby it
retains its antilinearity, and, due to a positive definite scalar product,
is also antiunitary. For clarity we denote this antiunitary charge
conjugation  by $\tilde C$. $\tilde C$ transforms $\tau $ into
$-\tau $ and $p_\mu $ into $-p_\mu $ with $x$ invariant.
 The parity transformation $P$ remains unitary and the
time inversion transformation $T$ antiunitary, since both can first be
defined on the particle sector alone and then carried over to the
antiparticle sector with help of the charge conjugation transformation. The
time-inversion carries $t$ into $-t$ and $\tau $ into $-\tau $ , so that
particles remain particles. The combined $\tilde CPT$-transformation, which
leaves $\tau $ invariant, is unitary. Since antiunitary transformations do
not have eigenvalues and eigenstates we cannot define $\tilde C$ or $\tilde
CP$ eigenstates in our theory, but we can define $\tilde CT$and $\tilde CPT$
eigenstates. The action of the symmetry transformations on spin 0 and spin
1/2 wave functions is identical with that of quantum mechanics. The
infinitesimal generator $i\partial _\tau $ of translations of the invariant
parameter is invariant under all these transformations, whereas the
infinitesimal generator $i\partial _t$ of time translations is invariant
only under $P$ and $T$, changing sign under $\tilde C$ and $\tilde CPT$. As
a consequence any Hamiltonian must be indefinite in a $\tilde CPT$-symmetric
theory.

We consider some relevant subsystem $\Psi _i$ of a physical system, with a
finite number $i=1,..,n$ of states, for which an effective evolution
equation is assumed. The development of $\Psi _i$ in time is governed by an
effective Hamiltonian matrix $\tilde H_{ij}$,
\begin{equation}
\label{105}i\partial _t\Psi _i=\tilde H_{ij}\Psi _j,
\end{equation}
whereas the evolution in proper time $\tau $ is described by an effective
mass matrix $M_{ij}$ with
\begin{equation}
\label{104}i\partial _\tau \Psi _i=M_{ij}\Psi _j.
\end{equation}
If the invariance of the subsystem under $\tilde CPT$ is assumed,
 the mass matrix is
invariant under $\tilde CPT$, as $i\partial _\tau $ is, whereas the
Hamiltonian $\tilde H_{ij}$ is transformed into $-\tilde H_{ij}$ under $%
\tilde CPT$, as is $i\partial _t\rightarrow -i\partial _t$. The mass matrix
may be choosen to be positive definite, the Hamiltonian necessarily is
indefinite.

\section{The neutral K-meson system}

For all definitions which are not standard we use a tilde.

The $K^0$ meson is a spin 0 particle consisting of a $d$ and a $\bar s$
quark, which have spin 1/2.  Since $\tilde CP$ is not unitary, we use $%
\tilde CPT$ to define the antiparticle to the $K^0$ meson, which we denote
by $\widetilde{\bar K}^0,$ by
\begin{equation}
\label{102}\tilde CPT\mid K^0\rangle =\mid \widetilde{\bar K}^0\rangle
,\quad \tilde CPT\mid \widetilde{\bar K}^0\rangle =\mid K^0\rangle ,
\end{equation}
so that we have $\tilde CPT$ eigenstates
\begin{equation}
\label{103}
\begin{array}{c}
\mid \tilde K_1\rangle =\frac 12\left( \mid K^0\rangle -\mid
\widetilde{\bar K}^0\rangle \right) ,\quad \tilde CPT=-1, \\ \mid \tilde
K_2\rangle =\frac 12\left( \mid K^0\rangle +\mid \widetilde{\bar K}^0\rangle
\right) ,\quad \tilde CPT=+1.
\end{array}
\end{equation}
We  describe the $K^0$ and $\ \widetilde{\bar K}^0$ mesons by a two-state
vector $\Psi =(\mid K^0\rangle ,\mid \widetilde{\bar K}^0\rangle )$, where $%
\tilde CPT$ interchanges both components through the multiplication by the $%
2\times 2$ matrix $\left(
\begin{array}{cc}
0 & 1 \\
1 & 0
\end{array}
\right) $. Assuming a $\tilde CPT$-invariant evolution equation the $2\times
2$ mass matrix is invariant under $\tilde CPT$,  which implies
\begin{equation}
\label{106}M_{11}=M_{22}\quad {\rm  and }\quad M_{12}=M_{21}.
\end{equation}
It follows that $\mid \tilde K_1\rangle $ and $\mid \tilde K_2\rangle $ are
eigenstates of $M$, just as in standard theory. Now in standard theory the
assumption that both Hamiltonian and mass matrix are invariant under the $CP$%
-transformation leads to the conclusion that $\mid \tilde K_1\rangle $ and $%
\mid \tilde K_2\rangle $ are also eigenstates of the Hamiltonian, and cannot
mix in time. The experimentally observed decay of the longest-living state
into both $CP$-eigenstates hence implies that the physical states $K_L$ and $%
K_S$ are distinct from $\mid K_1\rangle $ and $\mid K_2\rangle $ and thus $CP
$-symmetry must be violated \cite{13}. This is different in our theory. $%
\tilde CPT$-invariance in our theory implies that our Hamiltonian $\tilde H$
changes sign under $\tilde CPT$, so we have
\begin{equation}
\label{107}\tilde H_{11}=-\tilde H_{22}\quad {\rm  and }\quad
 \tilde H_{12}=-\tilde
H_{21}.
\end{equation}
Transforming $\tilde H$ into the basis $\Psi ^{\prime }=\left( \mid \tilde
K_1\rangle ,\mid \tilde K_2\rangle \right) $ yields
\begin{equation}
\label{108}\tilde H^{\prime }=\frac 12\left(
\begin{array}{cc}
1 & -1 \\
1 & 1
\end{array}
\right) \tilde H\left(
\begin{array}{cc}
1 & 1 \\
-1 & 1
\end{array}
\right) =\left(
\begin{array}{cc}
0 & \tilde H_{11}+\tilde H_{12} \\
\tilde H_{11}-\tilde H_{12} & 0
\end{array}
\right) .
\end{equation}
 The structure of $\tilde H^{\prime }$
is a general consequence of the Hamiltonian being $\tilde CPT$-odd. If $A$
and $B$ are any two $\tilde CPT$-eigenstates with eigenvalues $a,b\in
\left\{ +1,-1\right\} $ then
\begin{equation}
\label{108a}\langle A\mid \tilde H\mid B\rangle =\langle \tilde CPT\,A\mid
\tilde CPT\,\tilde H\left( \tilde CPT\right) ^{-1}\tilde CPT\mid B\rangle
=-ab\langle A\mid \tilde H\mid B\rangle
\end{equation}
so that only states with different $\tilde CPT$-eigenvalues yield nonzero
matrix elements.
{}From (\ref{108}) we see that the states $\mid \tilde K_1\rangle $ and $\mid
\tilde K_2\rangle $ always mix in time if $\tilde H$ is nonzero. The $\tilde
CPT$-eigenstates are not preserved in time, the eigenstates of the
Hamiltonian are not $\tilde CPT$-eigenstates. Hence, if we start with an
arbitrary initial state and wait until only the slowest-decaying eigenstate
has survived, we will always observe that this state decays into both
channels of $\tilde CPT$-eigenstates. So we are not forced to assume $\tilde
CPT$- or $\tilde CP$- violation in order to explain {\it qualitatively, }%
from first principles, the experimentally observed phenomenon of the
long-living $K_L$ decaying into different $\tilde CPT$-eigenstates.
The dynamical mixing of the $\mid \tilde K_1\rangle $ and $\mid \tilde
K_2\rangle $ also offers an explanation for the fact that $CP$-violation
practically is not seen in the decay of the physical $K_S$ \cite{13}:
The short decay
time provides no time for the mixing process to happen.
 A numerical study of a simple model
system confirms the mechanism as described \cite{Hannibal5}.


\begin{thebibliography}{99}
\bibitem{Bog}  {N. N. Bogolubov, A. A. Logunov, and I. T. Todorov, {\it %
Introduction to Axiomatic Quantum Field Theory}. (Benjamin, London, 1975)}

\bibitem{Streater}  {R. F. Streater and A. S. Wightman, {\it PCT, Spin \&
Statistics, and All That}. (Benjamin, New York, 1963)}

\bibitem{ReedSimon}  {M. Reed and B. Simon, {\it Methods of Modern
Mathematical Physics Vol II: Fourier Analysis, Self-Adjointness}. (Academic
Press, New York, 1975) }

\bibitem{Stueckelberg}  E. C. G. Stueckelberg,{\it \ Helv. Phys. Acta} {\bf %
14}, 322 and 588 (1941) and {\it Helv. Phys. Acta} {\bf 15} (1942) 23

\bibitem{Fanchi}  J. R. Fanchi, {\it Found. Phys.} {\bf 23}, 487 (1993); J.
R. Fanchi, {\it Parametrized Relativistic Quantum Theory} (Kluwer,
Dordrecht, 1993) and references therein

\bibitem{Feynman}  R. P. Feynman, {\it Phys. Rev.} {\bf 74}, 939 (1948) and
{\it Phys. Rev.} {\bf 80}, 440 (1950)

\bibitem{Hannibal1}  {L. Hannibal, {\it Int. J. Theor. Phys.} {\bf 30}, 1431
(1991)}

\bibitem{Hannibal2}  {L. Hannibal,{\it \ Int. J. Theor. Phys.} {\bf 30},
1445 (1991) }

\bibitem{Arshansky}  L. P. Horwitz and R. Arshansky, {\it J. Phys. A:\ Math.
Gen.} {\bf 15}, L659 (1982)

\bibitem{Wigner}  E. Wigner, {\it Ann. Math.} {\bf 40}, 149 (1949); M. A.
Naimark, {\it Linear Representations of the Lorentz Group} (Pergamon, New
York, 1964)

\bibitem{Hannibal5}  L. Hannibal, {\it Found. Phys. Lett.} {\bf 8}, 309 (1995)

\bibitem{Hannibal3}  L. Hannibal, {\it Rep. Math. Phys.} {\bf 33} 77, (1993)

\bibitem{Hannibal4}  L. Hannibal, {\it Found. Phys. Lett.} {\bf 7}, 551
(1994)

\bibitem{13}  J. H. Christensen, J. W. Cronin, V. L. Fitch and R. Turlay,
Phys. Rev. Lett. {\bf 13}, 138 (1964); L. K. Gibbons et al., Phys. Rev.
Lett. {\bf 70}, 1199 and 1203 (1993); Particle Data Group, L. Montanet et al.,
 Phys. Rev. D {\bf 50}, 1173 (1994) and references therein;
  T. T. Wu and C. N. Yang, {\it Phys. Rev. Lett.} {\bf 13}, 380
(1964); V. Barmin et al., Nucl. Phys. {\bf B247}, 293 (1984); L.
Wolfenstein, Ann. Rev. Nuc. Sci. {\bf 36}, 137 (1986);

\end{thebibliography}
\end{document}